\documentclass[final,3p,times,twocolumn,authoryear]{elsarticle_acc}
\usepackage{graphicx}
\usepackage{hyperref}
\usepackage[ansinew]{inputenc}
\usepackage{amsmath}
\usepackage{url}
\usepackage{color}
\usepackage{overpic}

\journal{Icarus}

\def\mms{$\mathrm{mm\,s^{-1}}$}
\def\cms{$\mathrm{cm\,s^{-1}}$}
\def\metersecond{$\ \mathrm{m\,s^{-1}}$}
\def\sio{$\mathrm{SiO_2}$}
\def\mum{$\mu$m}
\def\density{$\mathrm{kg\,m^{-3}}$}

\begin{document}
\begin{frontmatter}

\title{Free Collisions in a Microgravity Many-Particle Experiment. IV. - Three-Dimensional Analysis of Collision Properties}

\author{R. Weidling} \ead{r.weidling@tu-braunschweig.de}
\author{J. Blum}
\address{Institut für Geophysik und extraterrestrische Physik, Technische Universität Braunschweig,\\Mendelssohnstr. 3, D-38106 Braunschweig, Germany}

\begin{abstract}
The bouncing barrier, a parameter combination at which dust particles in the protoplanetary disk always rebound in mutual collisions, is one of the crucial steps of planet formation. In the past years, several experiments have been performed to determine the mass and velocity regimes at which perfect bouncing does occur and those where there is a chance of the aggregates sticking together. We conducted a microgravity experiment, which allows us to investigate free collisions of millimeter-sized \sio\ dust aggregates at the relevant velocities. We analyzed 52 collisions in detail with velocities of $3.4 \times 10^{-3}$\metersecond\ to $6.2 \times 10^{-2}$\metersecond\ and found four of them leading to sticking, while the other aggregates rebounded. Three out of the four sticking collisions occurred at velocities where previously only bouncing had been predicted. Although the probability for sticking is low, this opens a new possibility for growth beyond millimeter sizes. Our setup allowed us to obtain the complete three-dimensional collision information. Since most previous experiments were interpreted based on two-dimensional information, we compare our three-dimensional values with those obtained if only one projection had been available. We find that the error of a two-dimensional analysis of the collision velocity is very small. The distribution of the coefficient of restitution in the two-dimensional view is representative of the real case, but for any given collision its value can be far off. Impact parameters always have to be analyzed three-dimensionally, because the two-dimensional values are not meaningful in any way.
\end{abstract}


\end{frontmatter}

\section{Introduction}\label{sec:intro}
The early evolution towards planetesimals is dominated by the collisions among dust aggregates. Starting at sizes of about one micrometer, dust particles orbit around their young host star in a disk of gas and dust, called the protoplanetary disk (PPD). Several effects lead to relative velocities and, thus, to collisions among the particles, like Brownian motion, sedimentation and turbulence in the gas \citep{Weidenschilling:1977a}. A wealth of laboratory experiments has been conducted to investigate the possible outcomes of these collisions (see e.g. \citealp{BlumWurm:2008}, and \citealp{GuettlerEtal:2010}). While in general collisions at sufficiently low impact energies lead to sticking of the particles caused by van der Waals forces and those above a certain threshold lead to destruction, the details are more complex. \citet{GuettlerEtal:2010} compiled the experimental results into a first complete model for the collision behavior of dust aggregates. Based on previous experiments, this model predicts the outcome of a collision between two dust aggregates of given masses, relative velocity and porosity. The model has been used for local and 1D growth simulations in various PPD configurations by \citet{ZsomEtal:2010, ZsomEtal:2011}. They found that dust aggregates grow at most to sizes of about one centimeter. It should also be mentioned that the dust-aggregate collision model by \citet{GuettlerEtal:2010} used extrapolations for parameter-space regions where previously there had not been any experiments.

In order to experimentally fill these gaps in parameter space, we carried out a series of microgravity experiments at low collision velocities (see e.g. \citealp{WeidlingEtal:2012}, \citealp{KotheEtal:2013}, Brisset et al., \textit{submitted}). On the numerical side, the equations governing Smoluchovski solvers were improved \citep{WindmarkEtal:2012, WindmarkEtal:2012b, DrazkowskaEtal:2013} so that reliable simulations of dust-aggregate evolution in PPDs are possible. The simulations show that the introduction of particles with sizes beyond the bouncing barrier, which may form in dead zones or as CAIs, can grow towards planetesimal sizes due to collisions with mass transfer. An alternative avenue to planetesimals may be due to a succession of streaming and gravitational instability \citep{JohansenYoudin:2007, JohansenEtal:2007}. Local overdensities of cm-sized solid material in the gas disk attract more dust aggregates due to the streaming instability until the particle ensemble gravitationally collapses and forms bodies several kilometers in size. \citet{WahlbergJohansen:2014} show that for low-mass planetesimals the gravitational collapse is smooth so that the gravitationally infalling dust aggregates do not experience fragmentation. This assumption has been made by \citet{SkorovBlum:2012} in their model of comet nuclei and \citet{BlumEtal:2014} show that only this formation model is consistent with current comet activity. Thus, and regardless of the subsequent growth process, it is important to understand the details of the formation of cm-sized dust aggregates in PPDs.

In Section \ref{sec:experimental_setup} of this paper, we present an experimental setup providing free collisions among millimeter-sized dust aggregates. In Section \ref{sec:results}, we show the results of the experimental run and will discuss these in Section \ref{sec:discussion}. We will place the results in context and conclude in Section \ref{sec:conclusion}.

\section{Experimental Setup}\label{sec:experimental_setup}
The experimental setup consists of a cylindrical glass vacuum chamber filled with dust aggregates. The dust aggregates inside the glass tube are imaged with a high-speed camera at 500 frames per second. An LED array with diffusors behind the vacuum chamber provides homogeneous background illumination. The vacuum tube has a height of 50 mm and a diameter of 25 mm and is placed on top of a shaking mechanism. This mechanism consists of an eccentric wheel with the rotation axis being perpendicular to the chambers' symmetry axis. A more detailed description of the setup is given in \citet{WeidlingEtal:2012}. For the experiments we present in this paper, the setup called MEDEA-II was used in the Bremen drop tower, which is an upgraded version of the previously-flown MEDEA (Microgravity Experiment on Dust Environments in Astrophysics) experiment. The experiment was placed on the drop-tower catapult mechanism, allowing for a microgravity time of slightly more than 9 seconds, during which residual accelerations were negligible.

An equal-sided refraction prism with angles of 130$^{\circ}$ and $2\times$ 25$^{\circ}$ was mounted in the optical path of the camera in such a way as to yield two views of the glass vacuum chamber separated by a viewing angle of 30$^{\circ}$. This allows for the reconstruction of the three-dimensional trajectories of the particles while assuring that the dust aggregates visible in the two views can be correlated.

A turbomolecular pump was attached to the vacuum chamber to reduce the gas friction that was observed in \citet{WeidlingEtal:2012}. In the experimental run analyzed here, we evacuated the chamber to $3.5 \times 10^{-3}$ mbar. This is equivalent to a mean free path of the gas molecules of 28 mm. As this is much larger than the typical dust-aggregate diameter of about 1 mm, the particles are in the Epstein drag regime so that their stopping time $\tau_{F}$ can be calculated following \citet{Blum:2006}. For the experiment analyzed here, we get $\tau_{F}$ = 202 s. It should be mentioned that this is the e-folding time of the coupling of the dust-aggregate motion to the gas motion. Thus, a dust aggregate moving otherwise undisturbed in the vacuum chamber with the residual gas at rest loses at most $1-\exp\left( - 9 \mathrm{s/202 s} \right) = 0.043$ of their original velocity due to gas friction over the entire duration of the experiment. Since the parts of the aggregate trajectories analyzed here are much shorter than 9 s, we henceforth neglect the effect of gas drag on the motion of the dust aggregates.

During their preparation, the dust aggregates were sieved and only those remaining between two sieves with mesh sizes of 1.6 mm and 1 mm, respectively, were used. The sieved dust aggregates have an average mass of $7 \times 10^{-7}$ kg. They consist of irregular, polydisperse \sio\ with a material density of 2600 \density\ and 90\% of the monomer constituents possess a size between 0.1 to 10 \mum, rendering them ideal analog particles for PPD dust. As was shown in \citet{WeidlingEtal:2012}, the dust aggregates have a volume filling factor of 0.35.  More details on the dust properties can be found in \citet{BlumEtal:2006}.

Approximately 300 dust aggregates were placed into the vacuum chamber to obtain an average optical depth of around 0.3. This ensured a multitude of collisions between the dust aggregates while being transparent enough to track the trajectories of single particles. Assuming a mean velocity of 0.1 \metersecond\  the average collision time in this case is 0.16 s.

The shaking mechanism was set to three different frequencies during the course of the experiment. Each cycle lasted three seconds with the angular frequency $\omega$ being (in this order) 20.8 rad s$^{-1}$, 10.5 rad s$^{-1}$, and 4.9 rad s$^{-1}$, respectively. With an amplitude of the rotational motion of the shaker of $A=$ 0.5 mm, this results in a maximum wall speed during these phases of 9 \mms, 4 \mms, and 2 \mms, respectively, with a maximum centrifugal acceleration of 0.18 $\mathrm{m\,s^{-2}}$, 0.05 $\mathrm{m\,s^{-2}}$, and 0.01 $\mathrm{m\,s^{-2}}$, respectively. While collisions with the walls do happen, they do not alter the properties of the dust aggregates beyond what they already experienced from the preparation \citep{WeidlingEtal:2009}.

A movie of the entire experiment can be found in the online version of this article.

\section{Results}\label{sec:results}
\textit{Analysis.} In order to quantify the results of the multi-particle collision experiment, knowledge of the positions of all dust aggregates is required over the full duration of the experiment. Having obtained these, we can then calculate the instantaneous velocities of all dust aggregates. To reach this goal, we first binarized the images, separating the dark dust aggregates from the bright background. A threshold value for the intensity was selected in such a way as to preserve the projected area of the dust aggregates as quantitatively as possible. To rule out any adverse effects of the background illumination, which was not perfectly homogeneous, we compared the software-derived area of a sample of chosen dust aggregates with manual evaluations and found that the automated calculation yields very accurate values, independent of the position of the dust aggregate in the vacuum chamber and, therefore, the intensity of the local background illumination.

We then used the semi-automated tracking program briefly described in \citet{WeidlingEtal:2012}. The software stores the image number, position and cross-sectional area of each dust aggregate, under the condition that it possesses no overlap with any other particle. Additionally, it offers the possibility to store events, which are manual entries of collisions with other dust aggregates or the chamber wall. These events enable us to later sort through the tracks and quickly find the points of interest.

In a next step, we correlated the dust-aggregate events of interest in the two views of the experiment. This resulted in a total of 43 collisions in which both dust aggregates had been tracked sufficiently long to reconstruct their three-dimensional trajectories before and after the collision of the two dust aggregates. The trajectories were calculated by first linearly fitting the two 2D projections of each track separately. Then, the fit parameters were used for the calculation of the cartesian 3D trajectory properties. For simplicity, we chose for the left view of our three-dimensional images the coordinates $x$ and $z$, with $z$ being the coordinate along the symmetry axis of the glass cylinder and $x$ perpendicular to it and perpendicular to the viewing direction of the left image. The remaining $y$ coordinate in a cartesian coordinate system was then calculated from the coordinates of the center of mass of the considered dust aggregate in the left and right image, $x_\text{l}=x$ and $x_\text{r}$, respectively, each perpendicular to the symmetry axis of the glass cylinder and perpendicular to the respective viewing direction.
With an angle of 30$^{\circ}$ separating the two views, the $y$ coordinate is then given by $y=x_\text{r} / \sin 30^{\circ} - x_\text{l}/ \tan 30^{\circ} = 2 x_\text{r} - \sqrt{3}x_\text{l}$. To compensate for a possible offset of the two views due to inhomogeneities the optical system, we used for the $z$-coordinate the mean value of the respective coordinate in the two projections, $z_\text{l}$ and $z_\text{r}$, respectively, i.e. $z=0.5\cdot\left(z_\text{l}+z_\text{r}\right)$.

\textit{Comparison between 2D and 3D.} For free-floating particles in a closed system, one could expect an equipartition of energy in all translational directions. In that case, the 3D velocities are expected to be a factor of $\sqrt{3/2} = 1.22$ higher than the the projected 2D velocities. Here, however, the $z$-direction is also the direction of agitation, introducing the kinetic energy into the system preferably in this direction. Therefore, we expect a smaller difference than this factor 1.22 between 3D and 2D velocity. In \citet{WeidlingEtal:2012}, we determined a factor of 1.13 for experiments of the same type but with a different shaking profile. The exact difference between the velocity component in the direction of shaking, $v_\text{z}$, and the two components perpendicular to that, $v_\text{x}$ and $v_\text{y}$, respectively, depends on the shaking profile as well as on the number density of dust aggregates. In the experiment presented here, we found that the 3D velocities are on average a factor of 1.10 larger than the 2D velocities computed from the average of the two views. If we only compare the 2D velocities from the left or the right view with the 3D velocity, the ratio is a factor of 1.07 and 1.13, respectively. The median value for the ratio of 3D to 2D velocity is even less, only a factor of 1.02.

\begin{figure*}[tbh]
    \includegraphics[height=5.7cm, trim= 80 0 195 20, clip]{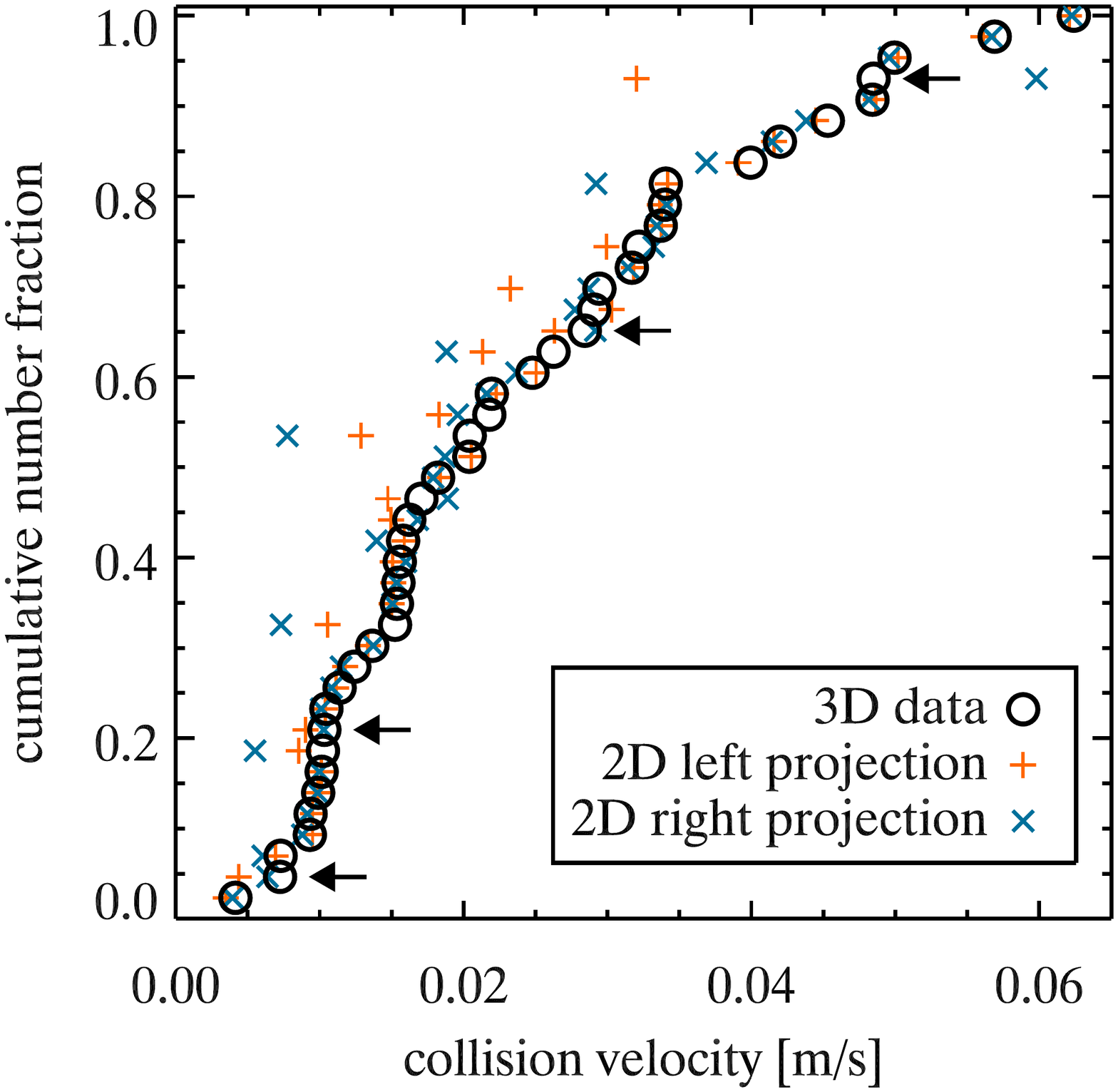}\includegraphics[height=5.7cm, trim= 150 0 195 20, clip]{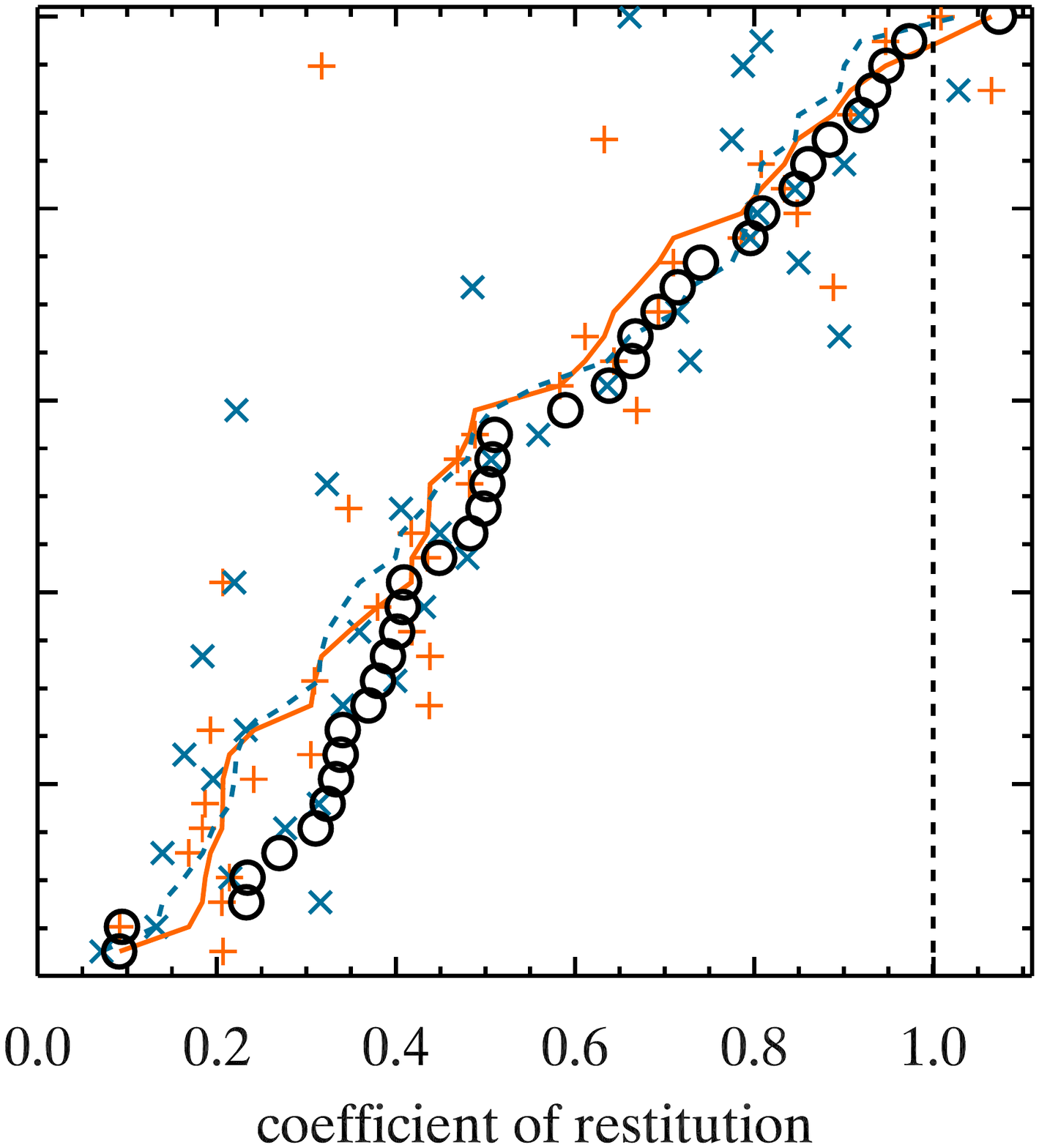}\includegraphics[height=5.7cm, trim= 150 0 100 20, clip]{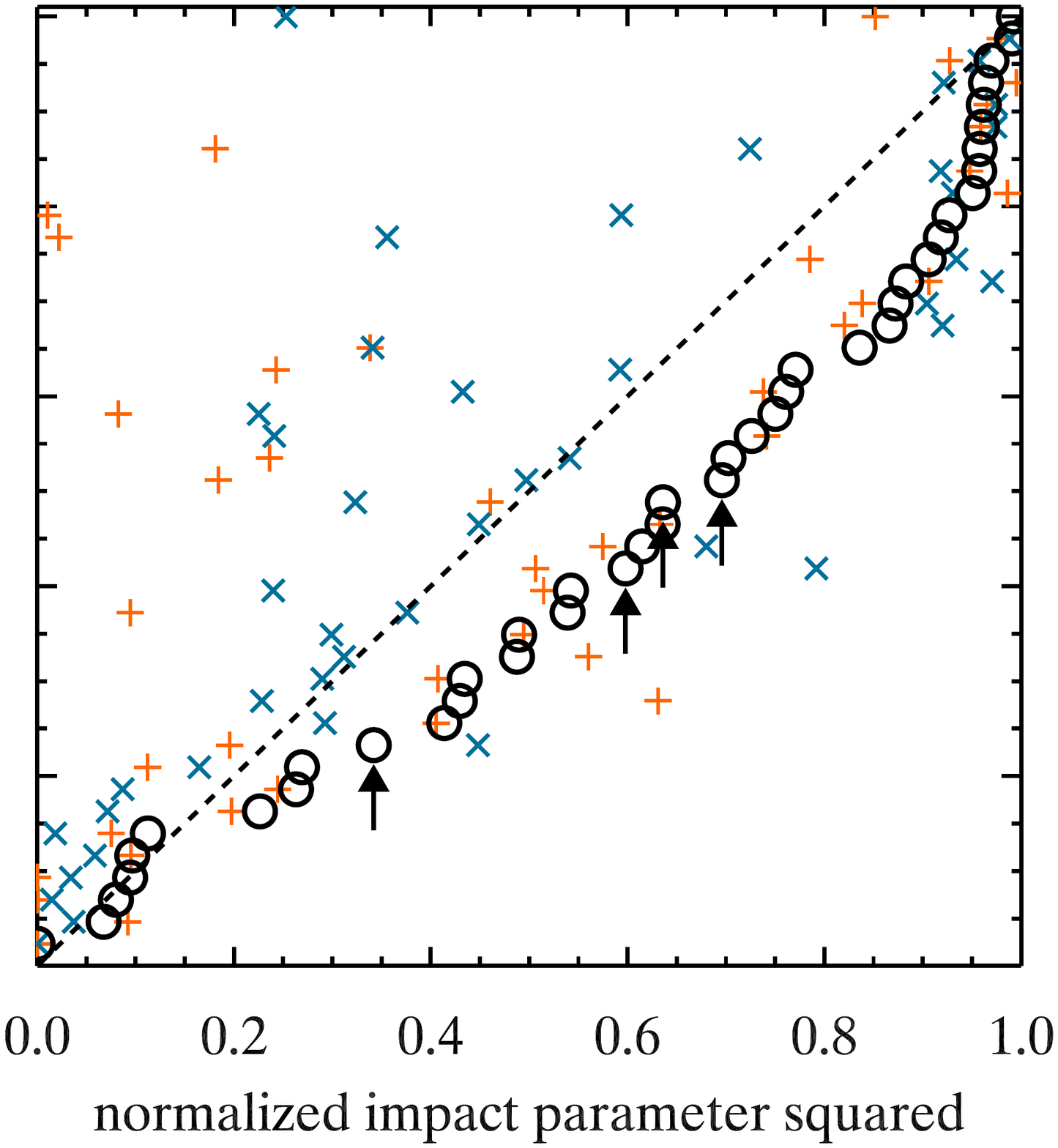}
    \caption{These plots show the comparison of the data based on the three-dimensional calculation (black circles) with those based on the two-dimensional projections (orange pluses and blue crosses). The collisions that result in sticking are marked by arrows. \textit{Left:} The collision velocities are in almost all cases very well represented by the 2D data. \textit{Middle:} The 2D coefficient of restitution can differ significantly from the true value, since the dust aggregates tend to possess a larger velocity component perpendicular to our field of view after a collision. The cumulative distributions of the 2D values for a larger number of collisions (orange solid and blue dashed lines) are slightly below the one for the 3D values. \textit{Right:} The 2D calculation of the impact parameter is often significantly underestimating the real value. Here, only a 3D analysis yields viable results. The dashed line represents the expected distribution of 3D impact parameters for arbitrary, isotropic collisions.}
    \label{fig:comparison2d3d}
\end{figure*}

Figure \ref{fig:comparison2d3d} (left) shows the collision velocity of the 43 pairs of dust aggregates for which we could analyze the trajectories three-dimensionally (black circles). For each collision, the orange pluses and blue crosses denote the values that were calculated based on the 2D projections only. As can be seen, the 2D information is in almost all cases a fair representation of the actual value. This means that the component of the velocity pointing in the direction of view is almost always small compared to the visible 2D velocity. Only in very few cases, like in the sticking collision (marked by an arrow) with the highest velocity, do the apparent velocities calculated based on the projections differ significantly. However, it should be noted that in this special case one dust aggregate was visible only for a very short time, leading to large uncertainties in the trajectory fits and, therefore, in the values of the velocity components.

The coefficient of restitution $\varepsilon$, the ratio of the relative velocity of two colliding dust aggregates after and before contact, was also calculated for the 2D and the 3D case (see Figure \ref{fig:comparison2d3d} (middle)). Here, $\varepsilon$ is an incomplete measure for the loss of energy in a collision, as for the moment we neglect rotational energy of the dust aggregates. Although we could clearly observe many of the aggregates rotating to a certain degree (see Section \ref{sec:discussion}), for most of them the images between two collisions were too few to yield a good estimate of the rotation frequency and rotational direction. As a result, it is possible to obtain values of $\varepsilon>1$ if rotational energy was transformed to translational energy in the collision; $\varepsilon>1$ may also occur if the viewing geometry leads to a more severe underestimation of the velocity before the collision than of the velocity after the contact.

On average, the three-dimensionally calculated coefficient of restitution is 20\% larger than its corresponding 2D values that are calculated based on the analysis of only one projection. The values for any specific collision can deviate significantly (see Figure \ref{fig:comparison2d3d} (middle)). Therefore, one cannot draw conclusions about the real 3D value if only one 2D value is known.The mean error of the 3D value is $\overline{\Delta\varepsilon} = \pm0.08$, the individual errors of the data points are shown in Figure \ref{fig:v_eps}. For most collisions the 2D values (orange pluses and blue crosses) underestimate the 3D value (black circles). Sorting the 2D values cumulatively (orange solid and blue dashed lines) reveals that the distributions in the 2D case are similar, but slightly smaller than the distribution of the 3D values. This is owing to the fact that after a collision the dust-aggregate velocities are more isotropic than before, when the velocities are determined by the container shaking.

For random collisions one would expect a uniform distribution of the normalized squared impact parameter $(b/R)^2$. Here, $R$ is the separation of the two centers of mass at the time of contact and $b$ is its projection perpendicular to the direction of the relative velocity. For this experiment, the values mostly follow the expected uniform distribution. Impact parameter values in the range $0.1 < (b/R)^2 < 0.4$ are slightly underrepresented statistically ($\sim$15\% instead of 30\% of all collisions) and  are slightly overrepresented statistically ($\sim$25\% instead of 10\% of all collisions) in the range $0.9 < (b/R)^2 < 1$. The statistics of the remaining values confirms the expectation, so all in all the distribution of impact parameters reasonably well describes random impacts. Having many collisions with a high impact parameter may be a selection effect of our analysis, since more central collisions (with a low $(b/R)^2$) yield a smaller coefficient of restitution (see Figure \ref{fig:bR_eps}), meaning that the dust aggregates are slowed down much more after a central collision so that one has to wait longer until the dust-aggregate projections are separated again. Due to the rather high optical depth in our experiment (see Section \ref{sec:experimental_setup}), it is, thus, much more likely that other particles' projections start to overlap with the dust aggregates of interest during this extended time interval, making it hard or even impossible to track the dust aggregates after the collision. In this case, we might still get information about the collision velocity and outcome, but for an analysis of the other parameters (e.g. impact parameter), we chose only those events for which we have the complete three-dimensional data, both from before and after the collision. Therefore, it is more likely that we could use the data obtained from a grazing collision with a high impact parameter, where the particle projections separate more quickly, than from a more central collision.

Whether the two-dimensional value is a good representation of the actual impact parameter strongly depends on the viewing geometry. Therefore, the only information that can be drawn from a two-dimensional value is the minimum normalized impact parameter in three dimensions. The real 3D value is always higher than or equal to the 2D value. Exceptions in the data, where some 2D values are higher than their corresponding 3D values, are due to uncertainties in the determination of the exact trajectories and may also result if one of the particles deviates significantly from the spherical shape.

\subsection{Collision parameters}\label{sub:collisionparameter}
Many collision models predict the rebound velocity for central collisions of two solid, spherical particles. They are based on material parameters, like, e.g., Young's modulus and specific surface energy, and yield the collision velocity below which the particles stick to each other and the coefficient of restitution for higher velocities. In \citet{WeidlingEtal:2012}, we showed that this concept can be transferred to dust-aggregate collisions used in the MEDEA setup. There, we followed the approach of \citet{ThorntonNing:1998} for adhesive elastic spheres, using a modified value for the surface energy to account for the structure of our porous dust aggregates to derive the sticking velocity of our aggregates.

\begin{figure}[tb!]
    \includegraphics[width=\columnwidth]{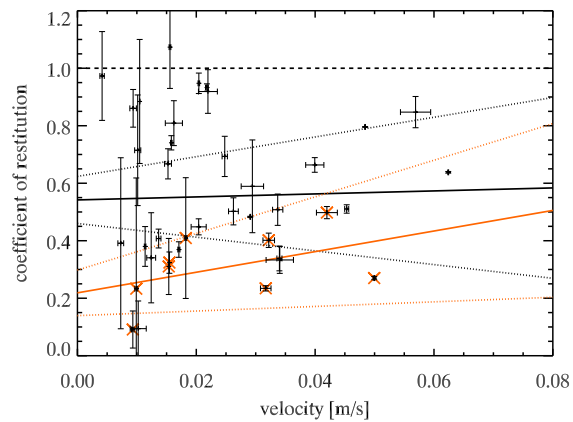}
    \caption{Coefficients of restitution of the three-dimensionally analyzed bouncing collisions as a function of collision velocity. Obviously, no clear trend can be made out; a linear fit (solid line) is shown for comparison. The $1\sigma$-uncertainties of the linear fit function are enveloped by the dotted lines and show that both an increase or a decrease of the coefficient of restitution with increasing velocity is plausible. The data that is additionally marked by an orange cross marks central collisions, where $(b/R)^2 < 0.4$. A linear fit of this subset shows the coefficient of restitution to grow with increasing velocity.}
    \label{fig:v_eps}
\end{figure}

However, in the experiment analyzed here, it is not possible to make any statement about the sticking velocity based on systematics of the coefficient of restitution data. We find that the coefficient of restitution possesses a mean value of $\varepsilon = 0.55$ and scatters widely from $\varepsilon = 0.09$ to $\varepsilon = 1.07$, while the collision velocities span only one order of magnitude (see Figure \ref{fig:v_eps}). The single value greater than 1 is an effect of rotational energy being transformed into translational kinetic energy in the collision. The errors in the velocities were derived from the $1\sigma$-uncertainties in the fits to the trajectories, the error in $\varepsilon$ by error propagation.

We can find no trend in our data for how the coefficient of restitution changes with velocity. Fitting the model by \citet{ThorntonNing:1998} directly to the data or to data bins did not yield any viable results. A linear relation between the coefficient of restitution and the impact velocity can also be ruled out, the Pearson correlation coefficient is only 0.001 (see Figure \ref{fig:v_eps}). The formal linear fit (solid black line) is nearly constant, but the $1\sigma$-uncertainties (area enveloped by the dotted lines) show that the trend might as well be increasing or decreasing, rendering it impossible to predict whether the collisions occur at velocities above or below the yield velocity. The yield velocity marks the onset of plastic deformation within the colliding dust aggregates and is close to the inflection point of the curve of the coefficient of restitution in the \citeauthor{ThorntonNing:1998} model. It is also important in order to determine the sticking velocity from the velocity dependence of the coefficient of restitution.

\begin{figure}[tb!]
    \includegraphics[width=\columnwidth]{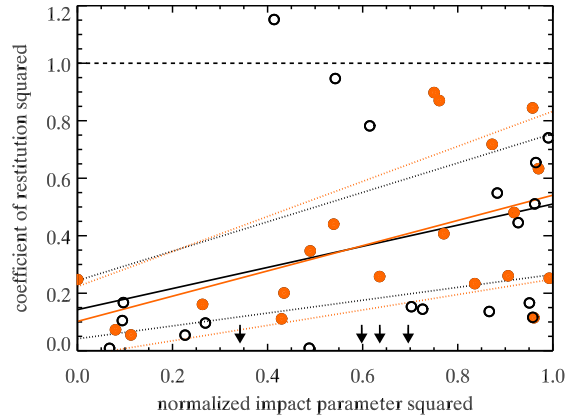}
    \caption{The squared coefficient of restitution $\varepsilon^2$ as a function of the squared normalized impact parameter $(b/R)^2$. Collisions with a velocity below 0.02 \metersecond\ are marked by open black circles, those with higher velocities with filled orange circles. A linear fit to all data (black solid line with the 1$\sigma$-range enveloped by the dotted lines) yields approximately the same values for central and grazing collisions as in \citet{BlumMuench:1993} and \citet{WeidlingEtal:2012}. Taking only collisions with a velocity above 0.02\metersecond\ into account (orange solid line) yields a similar result. The arrows mark the sticking collisions, which seem to show random impact parameters.}
    \label{fig:bR_eps}
\end{figure}

Since the impact parameter clearly has an influence on the coefficient of restitution (see Figure \ref{fig:bR_eps}), we investigated the result of central collisions, where $(b/R)^2 < 0.4$. This threshold was chosen since for higher impact parameters the scattering of the coefficient of restitution increased rapidly. The central collisions are marked by an additional orange cross in Figure \ref{fig:v_eps}. The orange solid line shows the result of a linear fit to this subset of data, with the dotted lines marking the $1\sigma$-uncertainties. The fit shows an increase of the coefficient of restitution with increasing velocity, suggesting that our velocity range is in between the sticking velocity and the yield velocity of the \citeauthor{ThorntonNing:1998} model. However, due to the limited number of central collisions, we can not quantify these two velocities.

Therefore, we think the coefficient of restitution in our experiment can best be described as a uniform distribution (see Fig. \ref{fig:comparison2d3d} (middle)) of $\varepsilon = 0.55 \pm 0.26$ over the given velocity range, with the latter being the standard deviation of the values. The scattering of the coefficient of restitution very likely is a result of randomly oriented collisions instead of head-on collisions (as modeled by \citet{ThorntonNing:1998}), the differences in particle size and, therefore, mass, and the irregularity of the surfaces of our dust aggregates, respectively. The considerable amount of scattering of the coefficient of restitution for collisions with velocities below 0.02 \metersecond\ can be attributed to rotation (see Section \ref{sec:discussion}). For these collisions the rotational energy cannot be neglected, however, we were not able to quantify the rotation for these dust aggregates. Depending on whether rotational energy was transferred into kinetic energy in the collision or vice versa, the coefficient of restitution can be dramatically increased or decreased. However, it should be noted that any real system with particles would probably also show scattering, which might lead to effects that are not expected based solely on idealized models.

The three-dimensional analysis also allows us to investigate a dependency of the coefficient of restitution on the impact parameter. In Figure \ref{fig:bR_eps} we show the squared coefficient of restitution $\varepsilon^2$ over the squared normalized impact parameter $(b/R)^2$. The circles represent the results of the bouncing collisions. Here, the orange filled circles show collisions with a collision velocity larger than 0.02\metersecond, where the scatter due to the importance of rotation is greatly reduced. Collisions slower than this are marked by the black open circles. The arrows show the impact parameters at which we observed sticking. The data suggest that the energy loss in a central collision is higher than for a grazing collision.

In \citet{WeidlingEtal:2012}, we observed the same effect with smaller dust aggregates of the same material colliding at approximately the same velocities. There, we found $\varepsilon^2 ((b/R)^2 = 0) = 0.12 \pm 0.04$ for central collisions and $\varepsilon^2 (1) = 0.51 \pm 0.12$ for grazing collisions from a linear least-squares fit to the data. \citet{BlumMuench:1993} found an identical trend in their experiments with dust aggregates made up of ZrSiO$_4$ and Aerosil. Their dust aggregates had about the same size as ours, but had lower volume filling factors and collided at higher velocities. In order to remove a possible effect of the velocity on the coefficient of restitution, they separated their results in velocity bins. In our case, we showed that there is no visible dependence between the two, so that we can use all the available data. \citet{BlumMuench:1993} found central collisions to lead to a squared coefficient of restitution of $\varepsilon^2 ((b/R)^2 = 0) = 0.04$ to $\varepsilon^2 ((b/R)^2 = 0) = 0.29$, depending on the material, and $\varepsilon^2 ((b/R)^2 = 1) = 0.50$ to $\varepsilon^2 ((b/R)^2 = 1) = 0.57$ for grazing collisions (their Table 2). All of the experiments agree very well with the theoretically derived value of $\varepsilon^2 ((b/R)^2 = 1) = 0.51$ by \citeauthor{BlumMuench:1993}.

Our results match the previous ones, with $\varepsilon^2 ((b/R)^2 = 0) = 0.14 \pm 0.10$ and $\varepsilon^2 (1) = 0.51 \pm 0.25$ (solid black line in Figure \ref{fig:bR_eps} with the dotted black lines enveloping the 1$\sigma$-range). This is the result of a linear least-squares fit to the data of all collisions. If we only take those values into account, where the collision velocity was larger than 0.2\metersecond\ (filled orange circles in \ref{fig:bR_eps}), the values change slightly to $\varepsilon^2 ((b/R)^2 = 0) = 0.10 \pm 0.12$ and $\varepsilon^2 (1) = 0.54 \pm 0.29$ (solid orange line with the 1$\sigma$-range given by the orange dotted lines). Since $\varepsilon^2$ is a measure of the energy loss in a collision, this means that about 40\% more of the initial kinetic energy is dissipated in central collisions, leading to more densely compacted aggregates.

Since we only observed four sticking collisions, we cannot be certain whether the impact parameter has an influence on the sticking probability or not. However, the spread of the values seems to indicate that sticking can occur independently of the impact angle.

\subsection{Collision outcomes}\label{sub:collisionresults}
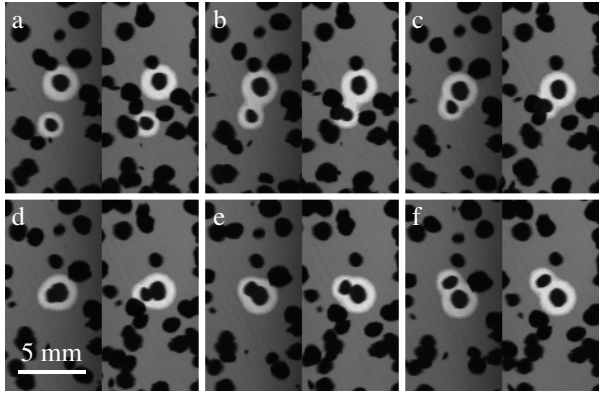
\begin{figure}[tb!]
    \begin{overpic}[width=\columnwidth]{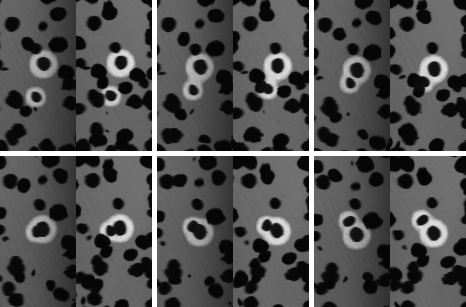}
    \put(1,61.5){\color{white} a}
    \put(35,61.5){\color{white} b}
    \put(68.5,61.5){\color{white} c}
    \put(1,28){\color{white} d}
    \put(35,28){\color{white} e}
    \put(68.5,28){\color{white} f}
    \put(2.5,5){\color{white} 5 mm}
    \linethickness{0.5 mm}\put(2,3){\color{white}\line(1,0){11.4}}
    \end{overpic}
    \caption{Image sequence showing a bouncing collision between two dust aggregates, highlighted by the lighter background. Subsequent images (a-f) possess a temporal spacing of 20 ms. Each image shows the region of interest in both projections (left and right part of the image). The dust aggregates collide with a relative velocity of $4.8\times10^{-2}$ \metersecond\ and visibly separate again after contact.}
    \label{fig:collision}
\end{figure}

In total, we observed 43 collisions for which we could obtain the complete three-dimensional trajectories of the involved dust aggregates. An example is shown in Figure \ref{fig:collision}, where a sequence of images shows the collision of two dust aggregates with a relative velocity of $4.8\times10^{-2}$ \metersecond\ from both available views. The aggregates are highlighted by the brighter background. They clearly separate again after contact.

 The observed dust aggregates collided with relative velocities between $4.1 \times 10^{-3}$\metersecond\ and $6.2 \times 10^{-2}$\metersecond . While the dust aggregates rebounded in 39 cases, we also observed them sticking together in 4 cases at velocities between $7.2 \times 10^{-3}$\metersecond\ and $4.8 \times 10^{-2}$\metersecond . As can be seen in Figure \ref{fig:comparison2d3d} the sticking collisions do not stand out in any way with respect to the collision velocity and impact parameter. Therefore, sticking of dust aggregates should be seen as a statistical process, most likely caused by minor differences in the structures of the dust aggregates. Examples of a bouncing and a sticking collision can be seen in the online version of this article.

An additional 9 collisions were analyzed that were visible in only one of the projections. In all of these latter cases, the dust aggregates bounced off each other. The collision velocities ranged from $3.4 \times 10^{-3}$\metersecond\ to $2.4 \times 10^{-2}$\metersecond .

\begin{figure}[tb!]
    \includegraphics[width=\columnwidth]{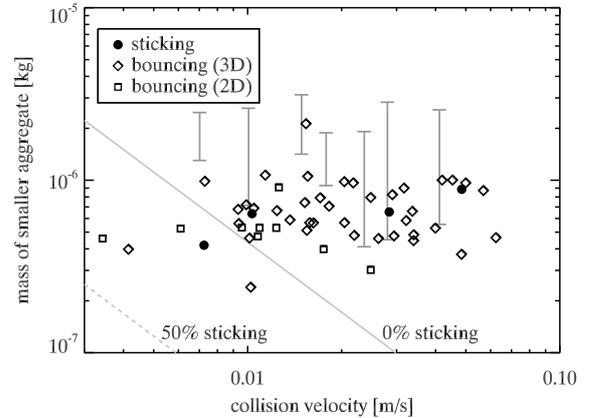}
    \caption{Collision outcome between dust aggregates. The diamonds and filled black circles denote the mass of the smaller of the two colliding dust aggregates and the three-dimensional collision velocity for the bouncing and sticking cases, respectively. The squares denote additional cases of bouncing dust aggregates for which only two-dimensional information was available. The lines denote the 50\% (dashed) and 0\% (solid) sticking probability after \citet{KotheEtal:2013}. The gray bars mark the bouncing collisions of monomers with dimers, with both the mass of the monomer (lower bound) and dimer (upper bound) marked.}
    \label{fig:model}
\end{figure}

In Figure \ref{fig:model} we show the results of all analyzed collisions with respect to their collision velocity and the mass of the smaller of the two dust aggregates, as was first introduced by \citet{GuettlerEtal:2010}. The gray solid line marks the combination of parameters where all collisions of faster or more massive dust aggregates would lead to bouncing according to \citet{KotheEtal:2013}. Slower or less massive dust aggregates may stick with a certain probability, with the 50\% level marked by the dashed line. Diamonds mark the bouncing collisions that we could analyze in 3D and squares those that were analyzed in 2D. The filled black circles denote the four sticking collisions.

It is worth noting that only one of the sticking collisions actually occurred in the parameter space for which the model derived in \citet{KotheEtal:2013} predicted sticking to be possible. The other three events indicate that the model needs to be updated yet again and that most likely the transition from perfect sticking to perfect bouncing collisions is even wider than in the current version of the model.

Additionally, we were able to deduce the collision velocities of 7 dust aggregates that collided with a dimer (gray bars in Figure \ref{fig:model} mark the mass of the monomer and the dimer, respectively). The velocities ranged from $7.0 \times 10^{-3}$\metersecond\ to $4.1 \times 10^{-2}$\metersecond . The orientations were random, the single dust aggregate sometimes collided with just one of the dust aggregates of the dimer and sometimes with both, in some cases along the axis of the dimer and in others perpendicular to it. Although we were able to distinguish the involved particles in both views, the collisions were calculated based on 2D information only, since analysis of the other view was not feasible in all these cases. One exemplary movie can be found in the online version of this article.

For the velocity of the dimer we tracked the center of mass of the projected area of the whole dimer, for the mass we assumed two dust aggregates of equal size making up the dimer. There were two additional collisions of this type, but in one case, the monomer was sticking to the wall and in the other case the dimer. Since these collisions were not free, we did not include them in our analysis. In all 7 cases of monomer-dimer collisions, the dimer survived the collision intact, meaning that the two dust aggregates making up the dimer clearly did not separate. On the other hand, none of the incoming monomer dust aggregates stuck to the dimer so that we did not observe any agglomeration beyond the dimer stage. Also, we could not observe how the dimers formed, as all of them emerged from dense regions where the overlap of the aggregate projections created a uniform dark area.

\section{Discussion}\label{sec:discussion}
\textit{Influence of electrical charges.} Since the van der Waals forces that may lead to dust aggregates sticking to each other are small, external influence is always a concern in experiments with dust analog material. Magnetic effects are ruled out by using non-magnetic \sio\ as the material of choice. However, electric charges and the corresponding Coulomb forces they exert are a concern, since the dust aggregates are insulators and charges could accumulate on their surface.

\begin{figure}[tb!]
    \includegraphics[width=\columnwidth]{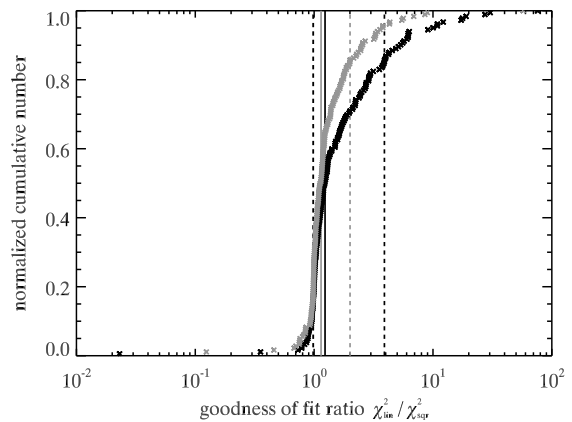}
    \caption{The ratio of the reduced goodness of fit for a linear and a quadratic fit function of all particles involved in a collision. Gray markings are used for the fits to the $x$-coordinates and black for the $y$-coordinates. The median values (solid black and gray lines) are only slightly larger than unity and the standard deviations (dashed black and gray lines) are within a factor of a few, indicating that the linear fits are not far worse than the square fits.}
    \label{fig:chisquare}
\end{figure}

\textit{Long-range influence.} Electrically charged dust aggregates do not move on temporally and spatially linear trajectories, but are accelerated in the vicinity of other charges. To test the hypothesis of charged dust aggregates in our experiment, we fitted a quadratic function to the trajectories and compared the results to the linear fits. For data following a linear trend, the reduced $\chi ^2$ of a linear, $\chi^2_{\text{lin}}$, and a square fit, $\chi^2_{\text{sqr}}$, are nearly identical, with the square one being slightly smaller. In Figure \ref{fig:chisquare} we show the goodness of fit ratio $\chi^2_{\text{lin}} / \chi^2_{\text{sqr}}$ for the tracks of all particles that were involved in collisions (gray for $x$-coordinates, black for $z$-coordinates), where the track consisted of at least 10 data points. This restriction was made since for very few data points, the trajectories may appear curved, due to binarization effects and rotation of the dust aggregates, while in reality they are not. The values within one standard deviation (dashed lines in Figure \ref{fig:chisquare}, note that the lower boundary is nearly identical in both coordinates) show that $\chi^2_{\text{lin}}$ is within a factor of 3 of $\chi^2_{\text{sqr}}$, meaning that the difference between the two kinds of fits is rather small. Therefore, we consider them to be of equal quality and the linear fit to be sufficient to describe the particles' trajectories so that the acceleration term is negligible. In those cases with a larger difference between the goodness of fit parameters, the trajectories may indeed be slightly curved, but these are only a few exceptions.

\textit{Short-range influence.} If we assume the dust aggregates to be charged and no charge to be transferred between the two particles in a collision, then the direction of the acceleration with respect to the direction of motion would change sign after the collision. While two aggregates with charges of opposite sign would be accelerated in the direction of motion before the collision, this would change after the collision when the particles separate again. If both aggregates have charges of equal sign the situation is reversed. Here, the acceleration points in opposite direction of the direction of motion before the collision and in the same direction afterwards.

However, an analysis of the formal accelerations (see above) showed that this reversal happens only in 5 out of 156 cases, if we take into account an angle of 30$^\circ$ of the direction of acceleration with respect to the direction of motion. On the other hand, for 12 dust aggregates the direction of the acceleration changed in the same way as the direction of motion. For the vast majority of the particles, the direction of motion and acceleration were not aligned before or after the collision. This is strong evidence that electric charges do not influence the dust-aggregate collisions in the experiment.

We also fitted a quadratic function to an increasing number of data points directly before a collision. If the particles are charged, Coulomb's law states that the force they exert on each other should increase with decreasing distance. However, there was no evidence for an increase in the acceleration in the fits with fewer data points or for sectors of the trajectory that were closer to the contact.

\textit{Charge estimate.} If we assume that the amount of acceleration that we did formally measure in the fit directly before a collision is solely generated by an electric field caused by charges on the dust aggregates, we can calculate the number of elementary charges on the dust aggregates. We will assume the particles to all have a mass of $7 \times 10^{-7}$ kg and to be equally charged. We will also assume that the dust aggregates consist of spherical monomers with a diameter of 1 \mum, which is between the median values for the mass and number weighted size distributions of the used dust (c.f. \citet{KotheEtal:2013}, their Figure 3).

For the acceleration, we used 0.02 $\mathrm{m\,s^{-2}}$, a typical value for the fit to the last 10 data points (20 milliseconds) before the collision. For the typical collision velocities of the dust aggregates (see Figure \ref{fig:comparison2d3d} (left)), this means that they were separated at most one diameter from the collision partner. To achieve this acceleration at this distance, $1.6 \times 10^6$ elementary charges on both dust aggregates would be necessary. This corresponds to about $3 \times 10^{-3}$ elementary charges per monomer within the dust aggregate or 3 elementary charges on each monomer on its surface. Even in this latter case, the additional attraction/repulsion due to the Coulomb force is many orders of magnitude smaller than the van der Waals attraction between two monomer grains in contact.

\begin{figure}[tb!]
    \includegraphics[width=\columnwidth]{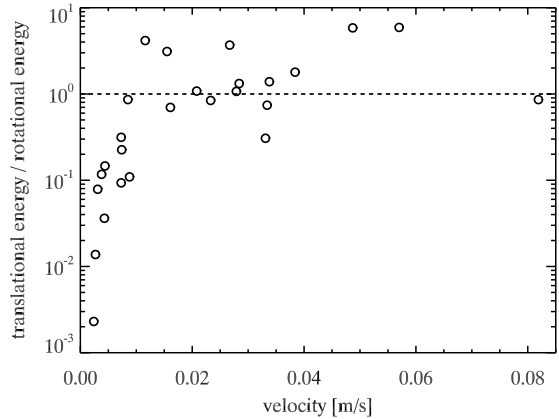}
    \caption{Ratio of translational and rotational kinetic energy of 26 dust aggregates as a function of their translational velocity. The horizontal dashed line marks equipartition of energy. Above this line, translation dominates rotation.}
    \label{fig:rotationalenergy}
\end{figure}

\textit{Rotational energy.} For 26 dust aggregates, we were able to determine the rotation frequency for parts of their trajectory. These dust aggregates are arbitrary tracked particles from both projections and random times of the experiment and were not involved in a collision during the time that was used to analyze their rotation. Due to the more ellipsoidal form of these dust aggregates, a rotation leads to a clear oscillation of the projected area of the dust aggregate. Fitting a sine function to the time series of the projected area yielded the rotation frequency which was found to be between 1.6 and 15.9 revolutions per second, independent of their velocity. However, if we compare the translational with the rotational energy, we find that for dust aggregates faster than a few centimeters per second, the translational kinetic energy on average is of the same order or even larger than the rotational energy (see Figure \ref{fig:rotationalenergy}, dashed line). The rotational energy, which was calculated assuming a spherical shape of the dust aggregates (ignoring the fact that in this case we would not have observed any oscillation of the projected area) dominates the particles' energy at low velocities. Calculations on the energy balance of these collisions, like the determination of the coefficient of restitution, ideally should take rotation into account. However, it is not always possible to determine the degree of rotation for a given dust aggregate in this kind of experiment, due to backlight illumination and the high optical depth.

While the colliding particles were rotating to a certain degree, we were not able to quantify the influence of the effect of rotation on the collision outcome. However, a comparison of the data in Figure \ref{fig:rotationalenergy} with those in Figure \ref{fig:v_eps} clearly shows that the spread in the values of the coefficient of restitution for a given velocity is visually much higher for impact velocities $< 2$ \cms than for higher velocities. These two velocity regions are exactly matching those for which rotation and translation dominate the total kinetic energy.

\textit{Dust-aggregate structure.} The internal structure of a dust aggregate can have a significant influence on the outcome of a collision. \citet{KotheEtal:2013} showed that clusters consisting of homogeneous sub-aggregates of about 150 \mum\ in diameter, stick at much higher velocities than homogeneous dust aggregates with the same mass. These dust aggregates consisted of monodisperse, spherical, micrometer-sized \sio\ particles. Brisset et al. (\textit{subm.}) observed the same effect for similar dust aggregates as well as for aggregates of identical size consisting of the same irregular, polydisperse, micrometer-sized \sio\ grains we used for our dust aggregates.

Thus, the exact properties of the monomer constituents do not seem to play a large role in these collisions, but the internal structure of a dust aggregate of a given mass does. Although we also observed dust aggregates of about $10^{-6}$ kg to stick at velocities higher than predicted by the model, the data by \citeauthor{KotheEtal:2013} and Brisset et al. (\textit{subm.}) suggest that the sticking probability for clusters of the same mass is even higher. This is likely due to the fact that their structure can be compacted much more easily, leading to a higher dissipation of energy in a collision.

Future revisions of the growth model will have to take the complete collision history of an aggregate into account in order to estimate the structure and, therefore, use the appropriate collision properties. If the growth of the dust aggregates in a PPD follows a very narrow size distribution, particles may very well be made up of smaller dust aggregates, which in turn might allow for growth beyond the maximum sizes found by \citet{ZsomEtal:2010}.

Another aspect that might favor further growth can be found in the result of the monomer-dimer collisions. For three of the observed dimers we cannot be certain how they were formed, since they became visible when the large cloud of particles at the beginning of the experiment became optically thin, so they might already have formed during preparation. The other four dimers, however, clearly formed during the remainder of the experiment duration. Thus, they formed at a collision velocity comparable to the subsequent collisions with the monomers. Therefore, although the chances of two dust aggregates to stick together at the velocities in the PPD are small, the resulting dimer survives subsequent collisions at similar velocities.  A few lucky winners might grow big enough in subsequent collisions to induce further growth by mass transfer as suggested by \citet{WindmarkEtal:2012}.

\section{Conclusions}\label{sec:conclusion}
As we have shown in Section \ref{sec:results}, for most of the collisions in our MEDEA-type experiments, the two-dimensional analysis of the velocity is reasonably accurate. The velocities obtained from one projected image may yield values that are on average about ten percent below the real value, but the overall picture is not affected by this. Since the velocity is the only parameter of a collision that is currently used for the collision model except for the mass, we can conclude that previous works yielded valid input. However, our data also show that the collision model has to be scrutinized, especially where transitions between regimes have not yet been backed by experimental data. The collisions analyzed in this paper reveal that the transition zone from perfect sticking to perfect bouncing is even broader than assumed previously.

The velocity regime of the MEDEA-type experiments with typical collision velocities of a few millimeters to a few centimeters per second is very relevant to the conditions in the solar nebula. However, it strongly depends on the particular nebula model and the amount of turbulence in which part of the disk these velocities occur for millimeter-sized particles. With an $\alpha$-parameter, which denotes the strength of turbulence, of $\alpha=10^{-3}$, the collision velocities for a dust-aggregate mass of $7 \times 10^{-7}$ kg at a stellar distance of 1 AU vary widely. For a high-density nebula \citep{Desch:2007}, the velocities are very low, only $1.9 \times 10^{-7}$ \metersecond. For a minimum mass solar nebula (MMSN) model \citep{HayashiEtal:1985, Weidenschilling:1977a}, the velocities at 1 AU are 0.2 \metersecond, close to those in our experiment. The highest velocities of 1.9 \metersecond\ are predicted for a low-density nebula \citep{BrauerEtal:2008a, AndrewsWilliams:2007}. \citet{JohansenEtal:2014} find velocities of 0.1 \metersecond\ for those collisions in the dead zone and 2 \metersecond\ with turbulence for a MMSN model. Therefore, for local simulations around 1 AU, our experiments are relevant for MMSN models. However, in global simulations, the conditions will always be met at some location in the disk. Also, the collisions will likely occur with a velocity distribution of some kind \citep{WindmarkEtal:2012b, GaraudEtal:2013}, which broadens the regions where velocities of a few millimeters per second to centimeters per second are relevant.

A three-dimensional analysis, like the one conducted here, is required to investigate the parameters describing the micro-physics of a collision. For the normalized impact parameter the real (3D) value may take any value between the number calculated for the two-dimensional projection and unity (see Figure \ref{fig:comparison2d3d} (right)). A two-dimensional analysis yields only a lower boundary to the real value, but the distribution of the 2D values should not be seen as representative of the real values.

The statistical distribution of the three-dimensional coefficients of restitution in this experiment is well represented by the two-dimensional values (see Figure \ref{fig:comparison2d3d} (middle)), but any single value may differ significantly. Therefore, any analysis of the coefficient of restitution should be based on a three-dimensional analysis to ensure that each value is correct. Since the collisions are random, looking at them from a fixed perspective also yields a random underestimation of the velocities before and after the collision. Therefore, a sufficient number of collisions may yield an estimate of the distribution of coefficients of restitution, but should not be treated as showing the whole picture.

The existence of "lucky winners" that may be necessary for growth beyond centimeter sizes \citep{WindmarkEtal:2012, DrazkowskaEtal:2013} is plausible from our experiments. However, statistics are limited in our experiment. \citet{KellingEtal:2014} analyzed tens of thousands of collisions and found all dimers to be destroyed again. The collisions in our experiment were completely free, whereas aggregates in the setup of \citeauthor{KellingEtal:2014} were subject to external forces. "Lucky winners" would need to have a smaller velocity than in our experiment or be among the few percent for which sticking occurs, despite having rather high velocities, but if they do form dimers, these are stable enough to survive subsequent collisions. Therefore, compound aggregates might grow further in future collisions until they are large enough to start growing by mass transfer instead of direct sticking. Only very few particles will reach these sizes, but the simulations showed that even a tiny fraction of larger particles is sufficient to quickly form bodies a few hundred meters in size.

\subsection*{Acknowledgements}
We are grateful to the Deutsches Zentrum für Luft- und Raumfahrt (DLR) for funding the experiments and drop tower campaigns under grants 50WM0936 and 50WM1236. We thank Stefan Kothe for valuable discussions.

\bibliographystyle{apalike}
\bibliography{literatur}

\end{document}